\documentclass[journal,onecolumn,12pt,nofonttune]{IEEEtran}
\ifCLASSINFOpdf
\else
\fi

\usepackage{graphicx}
\usepackage{color}
\usepackage{placeins}
\usepackage{float}
\usepackage{hyperref}
\usepackage{tabularx,colortbl}

\usepackage{amsmath,amsfonts,amssymb}
\usepackage{mathdots}
\usepackage{bbm}
\usepackage{mathtools, cuted}
\usepackage{amssymb,bm}
\usepackage{hyperref}

\usepackage{mwe}
\usepackage{overpic}
\usepackage{tikz}
\usetikzlibrary{calc,positioning}
\usepackage{array}
\usepackage{tabularx}
\usepackage[usestackEOL]{stackengine}
\usepackage{blindtext}
\usepackage{float}
\usepackage{siunitx}
\usepackage[english]{babel}
\usepackage{lipsum}  
\usepackage{booktabs}

\newcommand{\htext}[1]{%
	\makebox[0pt]{\Centerstack{#1}}
}

\newcommand{\vtext}[1]{%
	\makebox[0pt]{\rotatebox[origin=c]{90}{\Centerstack{#1}}}
}
\newcommand{\vtextf}[1]{%
	\makebox[0pt]{\rotatebox[origin=c]{-90}{\Centerstack{#1}}}
}

\newcommand{\chia}[2]{\chi_\text{#1}^{#2}}

\definecolor{amber}{rgb}{0.8, 0.28, 0.0}
\definecolor{ceruleanblue}{rgb}{0.16, 0.28, 0.75}
\definecolor{ao}{rgb}{0.0, 0.5, 0.0}

\newcommand{\sg}[1]{{\leavevmode#1}}
\newcommand{\tjs}[1]{{\leavevmode#1}}
\newcommand{\vt}[1]{{\leavevmode#1}}
\newcommand{\vtb}[1]{{\leavevmode#1}}
\newcommand{\kw}[1]{{\leavevmode#1}}

\makeatletter
\let\MYcaption\@makecaption
\makeatother
\usepackage[labelformat=simple,font=footnotesize]{subcaption}

\makeatletter
\let\@makecaption\MYcaption
\makeatother

\begin{document}
%
\title{\sg{Metasurface Near-field Measurements with Incident Field Reconstruction using a Single Horn Antenna}}

%
%
%

\author{Ville Tiukuvaara,~\IEEEmembership{Student Member,~IEEE,}
        Kan Wang,
        Tom J. Smy \\ and Shulabh Gupta,~\IEEEmembership{Senior Member,~IEEE}
\thanks{  Ville Tiukuvaara, Kan Wang, Tom J. Smy, and Shulabh Gupta are with Carleton University, Ottawa, Canada (e-mail: villetiukuvaara@cmail.carleton.ca). }
}

%
%

\markboth{}{}
%



\maketitle

\begin{abstract}
A simple method of superimposing multiple near field scans using a single horn antenna in different configurations to characterize a planar electromagnetic metasurface is proposed and numerically demonstrated. It can be used to construct incident fields for which the metasurface is originally designed for, which may otherwise be difficult or not possible to achieve in practice. While this method involves additional effort by requiring multiple scans, it also provides flexibility for the incident field to be generated, simply by changing the objective of a numerical optimization which is used to find the required horn configurations for the different experiments. The proposed method is applicable to all linear time-invariant metasurfaces including space-time modulated structures.  
\end{abstract}




%
\IEEEpeerreviewmaketitle

%
%
%
%
\IEEEPARstart{M}{etasurfaces} \sg{(MSs)} are the 2-D equivalent of metamaterials, the latter being a class of artificial engineered materials exhibiting peculiar electromagnetic properties \cite{Kuester:2003aa,caloz_electromagnetic_2005}. Generally constructed as arrays of deeply sub-wavelength resonant particles on a substrate, the geometry of the particles can be carefully designed to produce transformations of incident waves, including control of phase, amplitude, polarization, and direction of propagation. Recently, a major research direction has been ``intelligent'' MSs,  where the wave transformation can be electrically controlled, which could be a viable means of achieving in \sg{5G and future} wireless communications, the goal of manipulating and optimizing the propagation environment \cite{basar_wireless_2019} \sg{or to even create sophisticated illusions and holograms on the fly} \vt{\cite{smy_surface_2020}}. Other topics that have recently been studied are surfaces with time-varying properties \cite{caloz_spacetime_2020} and surfaces composed of particles with multipolar moments \cite{achouri_multipolar_2021}.  

\begin{figure}[h]
    \centering
    \subcaptionbox{Schematic\ \\\ \\}{
    \begin{overpic}[scale=1.1,grid=false,trim={0cm 0cm 0cm 0cm},clip]{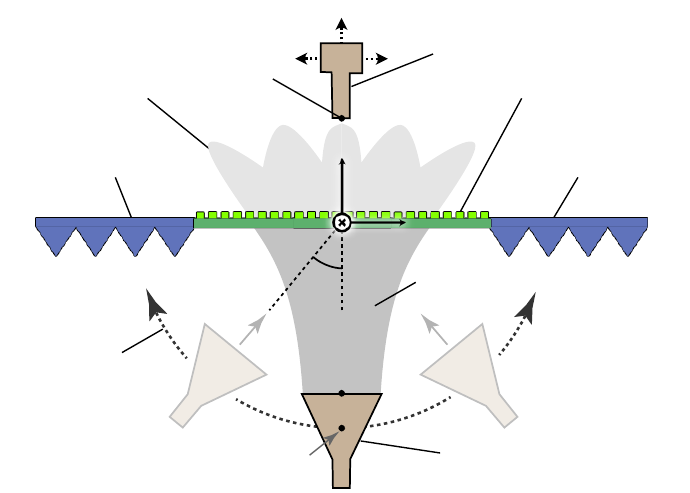}
        \put(36, 3){\htext{\scriptsize{}$(x_\text{bw},z_\text{bw})$}}
        \put(49.5, 17){\htext{\scriptsize{}$(x_\text{ap},z_\text{ap})$}}
        \put(33,62){\htext{\scriptsize{}$(x_\text{rx},z_\text{rx})$}}
        \put(64,0){\scriptsize{} \shortstack{Rectangular horn \\ antenna (Tx)}}
        \put(61,30){\scriptsize{}$\textbf{E}_\text{i}(x,z)$}
        \put(77,58){\scriptsize{}Metasurface}
        \put(15,58){\htext{\scriptsize{}$\textbf{E}_\text{t}(x,z)$}}
        \put(10,47){\htext{\scriptsize{}Absorbing material}}
        \put(90,47){\htext{\scriptsize{}Absorbing material}}
        \put(70,68){\htext{\scriptsize{}Probe antenna (Rx)}}
        \put(70,65){\htext{\tiny{}(scans $xz$ plane)}}
        \put(44, 29){\scriptsize{} $\theta$}
        \put(3, 13){\scriptsize{} \shortstack{Flexible \\ Positioning}}
        \put(58,36){\htext{\tiny{}$x$}}
        \put(48,49){\htext{\tiny{}$z$}}
    \end{overpic}
    }\quad
    \subcaptionbox{Near-field system at Carleton University}{
        \begin{overpic}[width=0.33\columnwidth,grid=false,trim={0cm 0cm 0cm 0cm},clip]{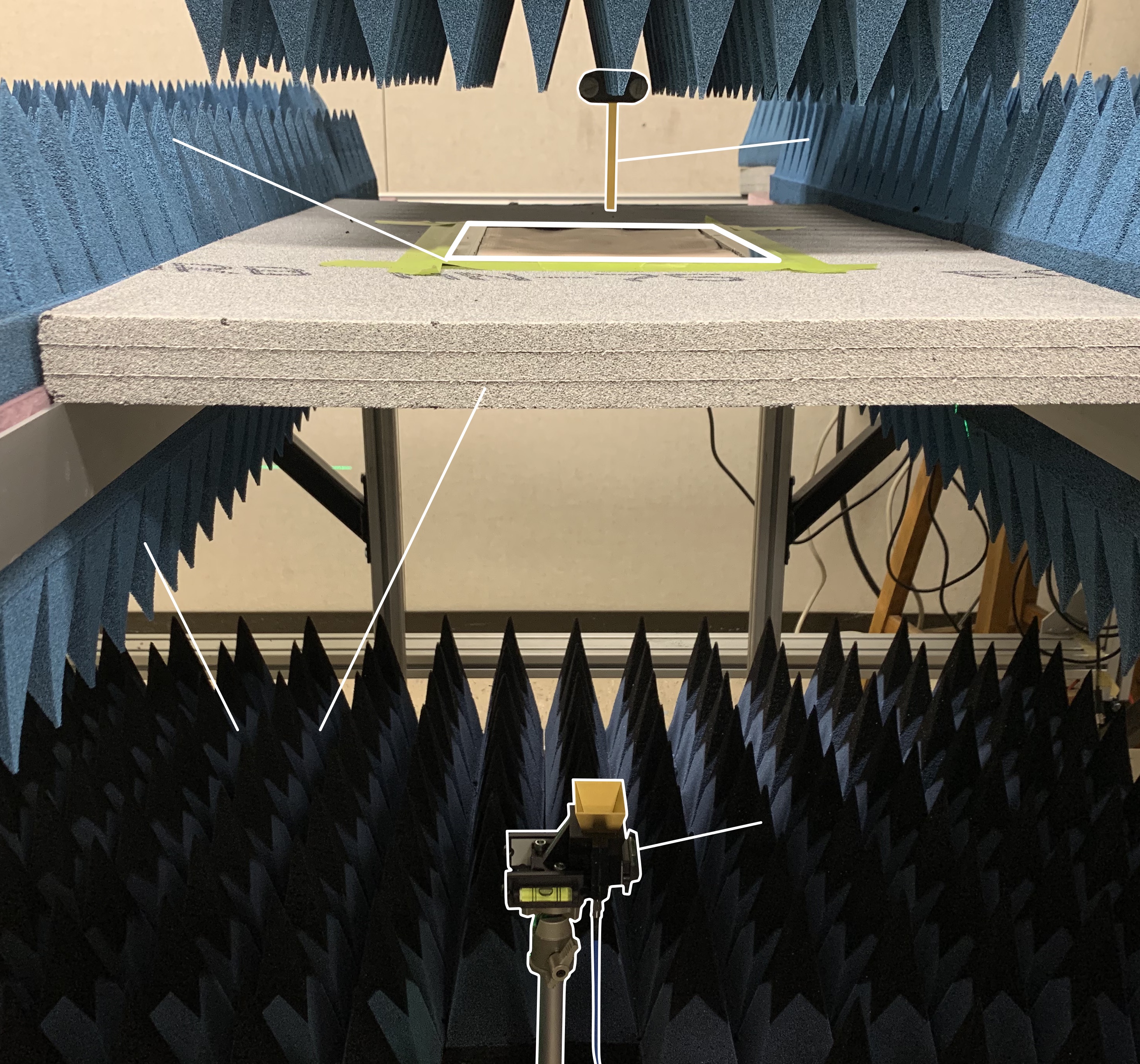}
            \put(64,18){\scriptsize{} \color{white}\shortstack{Rectangular horn\\antenna (Tx)}}
            \put(16,19){\scriptsize{} \color{white}\shortstack{Absorbing\\material}}
            \put(12,82){\scriptsize{} \color{white}\htext{Metasurface}}
            \put(71,77){\scriptsize{} \color{white}\shortstack{Probe\\antenna (Rx)}}
        \end{overpic}
    }
    \caption{\vtb{Near-field system for transmissive metasurface characterization.}}
    \label{fig:schematic}
\end{figure}
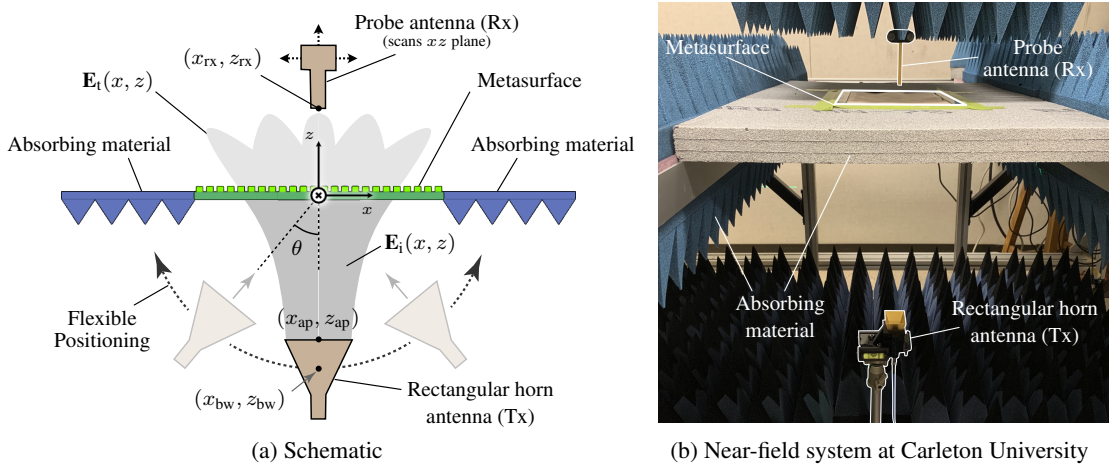

Such MSs are generally designed with a combination of simulations and models such as equivalent circuits or surface susceptibilities. These allow the designer to quickly calculate the scattered electric fields, given an incident field. Following fabrication, there are multiple approaches to the characterization of the MS, depending on the quantities which are desired. If the scattered far-fields are desired, an angular scan can be performed in an anechoic chamber. On the other hand, if the \sg{complex} electric field distribution next to the metsurface is desired, it can be probed using a near-field scanning system, as depicted in Fig.~\ref{fig:schematic}a \cite{yaghjian_overview_1986}. In the context of MSs, such a system works by illuminating the MS with an incident field $\textbf{E}_\text{i}(x,z)$ generated with a fixed antenna (Tx) with its aperture placed at $(x_\text{ap},z_\text{ap})$. The transmitted fields next to the MS, $\textbf{E}_\text{t}(x,z)$, are probed with a waveguide probe antenna (Rx) which can perform a scan along a line, plane, or volume and record the field at a collection of points $(x_\text{rx},z_\text{rx})$. Absorbing material is installed to prevent unwanted reflections for an accurate characterization of the MS, and can also be installed adjacent to the surface to eliminate diffraction around the edges, as depicted in Fig.~\ref{fig:schematic}a. \vt{An example of such a system used by the Metamaterials and Antennas Research Group (MARS) at Carleton University is in Fig.~\ref{fig:schematic}b,} \vtb{operating from \SIrange{26.5}{40}{GHz} with Eravent antenna models SAR-2013-28-S2 and SAP-28-R2 for Tx and Rx, respectively.} \sg{Compared to the standard far-field characterization, the prime benefit of a near-field system is its compact size and capability of measuring the detailed complex transmittance of the surface. The near-fields nevertheless can be used to compute the far-fields using standard near-to-farfield transformation procedures} \vt{\cite{yaghjian_overview_1986}}.

\sg{However,} it is quickly apparent that this system does not provide much flexibility for controlling the incident field. The designer is limited to using \tjs{the} specific Tx antennas which are available to them, and configuring the orientation and position of the Tx antenna. \sg{This is dominantly due to expensive horn antennas used in these systems in practice, where usage of multiple Tx antennas is not always possible and is not cost-effective.} Thus, it may be that it is not possible to experimentally produce the \sg{exact} incident fields that were used in simulation, and for which the MS is \sg{originally} designed. For example, consider a MS that was designed for a normally incident plane wave. Using a rectangular horn antenna for illumination, it is not possible to produce an ideal plane wave, \kw{\tjs{with} both uniform phase and amplitude\tjs{,}} as we will show. \sg{While }it is possible to approximate a plane wave by moving the horn antenna far away from the MS, this comes at the expense of losing much of the incident power, affecting the signal-to-noise ratio \sg{and \tjs{introducing}  undesired effects due to non-uniform phase distribution across the MS}. One approach which has been taken specifically to generate a flat phase is the use of a lens placed between the illuminating horn and the MS \cite{gagnon_material_2002,wong_polarization_2015,selvanayagam_design_2016}. This is based on the \textit{quasi-optical} approximation of the field generated by the horn being a Gaussian beam \cite{goldsmith_quasi-optical_1992}. In this case, the system can be analyzed within the framework of paraxial optics which can be used to design a lens which produces a beam waist (and hence a constant phase) at the location of the MS \cite{bruce_abcd_2006}. \sg{However, the typical spot-size generated using this lensed system is small, of the order of few centimeters for a typical Ka-band system, for instance, which is not sufficient to characterize larger size MSs (typically several tens of wavelengths), beyond which the phase flatness is significantly degraded. This greatly limits the physical area that can be field scanned.} In addition, a quasi-optical approach of modeling the horn field as a Gaussian beam reveals the \tjs{inherent trade-off present -- as the} phase and the amplitude uniformity cannot be optimized at the same time. One can form this conclusion directly from the \tjs{formulation for the Gaussian beam which has a uniform phase profile at the waist where the spot size is smallest and the magnitude variation sharpest.}

In this paper, we propose a \sg{novel} technique \sg{based on just a single Tx antenna}, which does not require additional components such as lenses, and provides flexibility in shaping the incident field \sg{including a flat uniform phase across a large  physical area}. The method involves multiple separate experiments with different incident and scattered fields, which are subsequently combined using superposition to produce the desired incident and scattered fields. \vtb{The application of the superposition principle assumes a linear system, which is the case for most MSs, including both linear time-invariant (LTI) and linear time-variant (LTV) MSs (non-linear MSs are notably excepted).} %
\sg{In this work,} we will numerically demonstrate the method using an integral equation (IE) simulator \cite{smy_ie-gstc_2021}, while the same procedure can be carried out in a laboratory setting.

The paper is structured as follows. In Section~\ref{sec:horn}, we show how the field generated by a rectangular horn can be modelled as a Gaussian beam, which provides a convenient analytical model. Using this model, in Section~\ref{sec:superposition} we show how a particular incident field---we use an example of a normally-incident plane wave---can be generated by the fields from multiple horns in different configurations. Finally, in Section~\ref{sec:parabolic-example} we apply this incident field to a MS which has been designed for a normally-incident plane wave, showing that it produces the correct scattered field while illumination with a single horn does not.

\section{Practical Metasurface Illumination}\label{sec:horn}
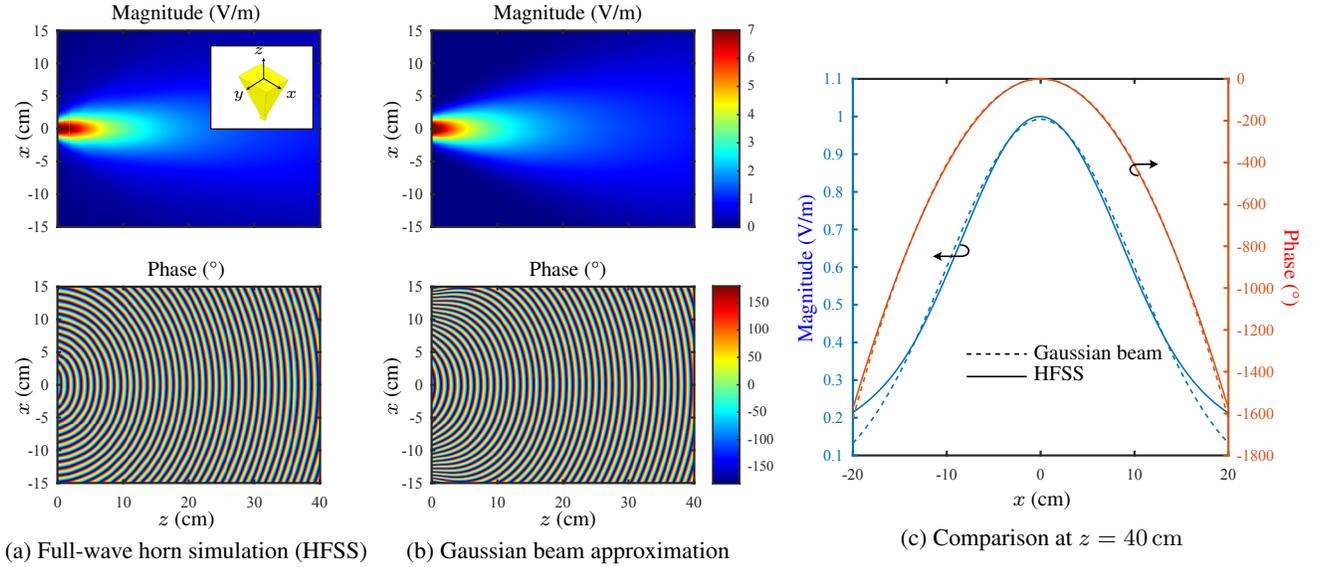
\begin{figure}[h]
    \centering
    \begin{overpic}[scale=1,grid=false,trim={0cm 0cm 0cm 0cm},clip]{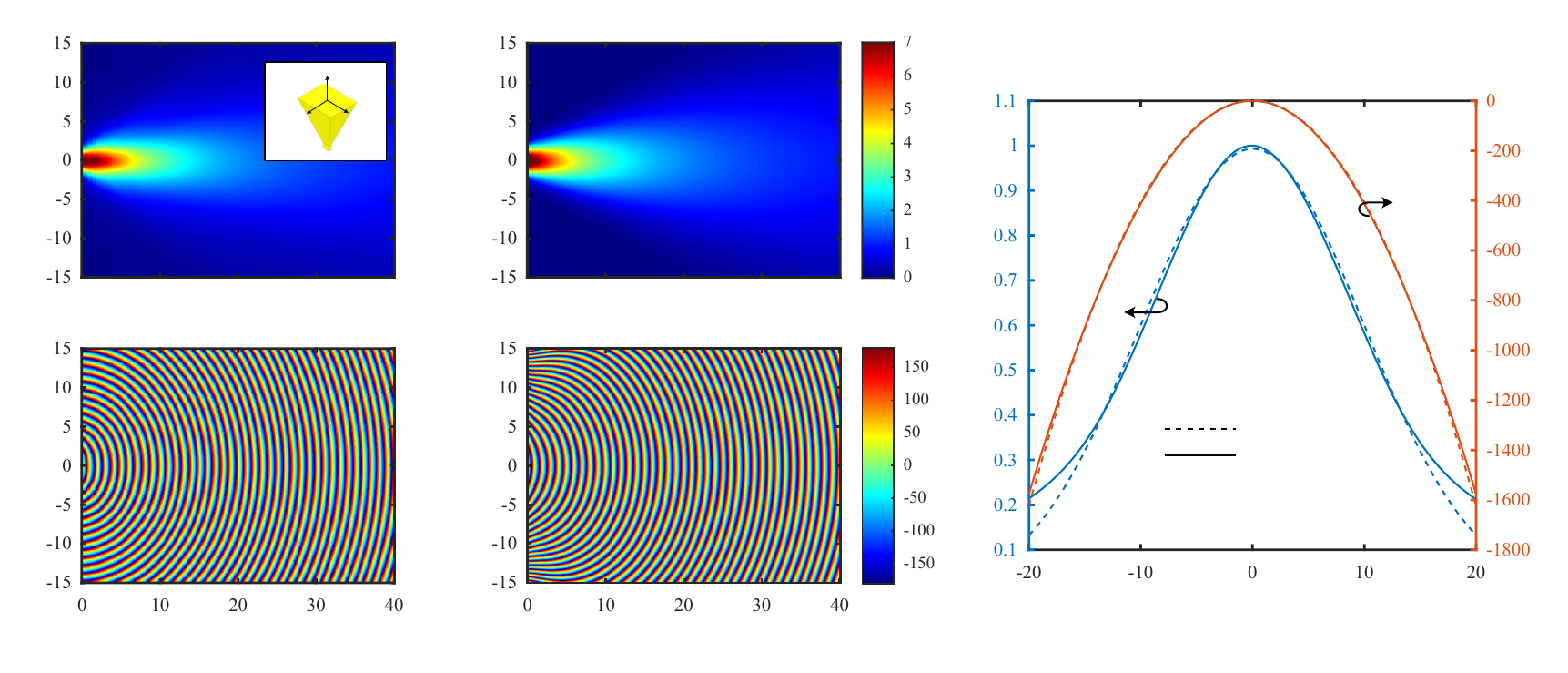}
		\put(15,2.5){\htext{\scriptsize{}$z$ (cm)}}
		\put(44,2.5){\htext{\scriptsize{}$z$ (cm)}}
		\put(2.5,12.5){\vtext{\scriptsize{}$x$ (cm)}}
		\put(2.5,32){\vtext{\scriptsize{}$x$ (cm)}}
		\put(30.5,12.5){\vtext{\scriptsize{}$x$ (cm)}}
		\put(30.5,32){\vtext{\scriptsize{}$x$ (cm)}}
		\put(15,0){\htext{\footnotesize{}(a) Full-wave horn simulation (HFSS)}}
		\put(15,21.5){\htext{\scriptsize{}Phase (\si{\degree})}}
		\put(15,41){\htext{\scriptsize{}Magnitude (V/m)}}
		
		\put(44,0){\htext{\footnotesize{}(b) Gaussian beam approximation}}
		\put(44,21.5){\htext{\scriptsize{}Phase (\si{\degree})}}
		\put(44,41){\htext{\scriptsize{}Magnitude (V/m)}}
		\put(80,1){\htext{\footnotesize{}(c) Comparison at $z=\SI{40}{cm}$}}
		\put(80,4){\htext{\scriptsize{}$x$ (cm)}}
		\put(62,21.5){\vtext{\scriptsize{}\color{blue} Magnitude (V/m)}}
		\put(99,21.5){\vtextf{\scriptsize{}\color{red} Phase (\si{\degree})}}
		\put(79.4,15.1){\scriptsize{}Gaussian beam}
		\put(79.4,13.5){\scriptsize{}HFSS}
		\put(19,35){\htext{\tiny{}$y$}}
		\put(23,35){\htext{\tiny{}$x$}}
		\put(20.5,38.3){\htext{\tiny{}$z$}}
	\end{overpic}
    \caption{The Eravant SAR-2013-28-S2 horn antenna was simulated using HFSS, and is well-approximated using a Gaussian beam having the parameters $w_0=\SI{1.00}{mm}$, $x_\text{bw}=\SI{0}{cm}$, and $z_\text{bw}=-\SI{4.14}{cm}$.}
    \label{fig:horn}
\end{figure}

We will consider the incident fields generated by the Eravant model SAR-2013-28-S2 rectangular horn antenna, which functions in the \SIrange{26.5}{40}{GHz} band. Fig.~\ref{fig:horn}a shows the electric field profile in the $H$ plane \vtb{($x-z$ plane)}, simulated using the full-wave Ansys HFSS simulator \vtb{at $f=\SI{30}{GHz}$} (simulation model inset in top right). For the simulation, the aperture of the horn is placed at $(x_\text{ap},z_\text{ap})=(0,0)$. We see that the horn produces curved phase-fronts, with the fields plotted along an observation line at $z=\SI{40}{cm}$ in Fig.~\ref{fig:horn}c for closer examination. There is significant curvature of the phase: along the observation line at a distance of $x=\SI{7.5}{cm}$ from the peak, the phase has decreased by \SI{230}{\degree} while the amplitude has reduced to 72\%. \sg{This is a significant deviation from a flat phase front of a uniform plane-wave, for instance, which is typically used in various MS designs.}

For the method we will present, \sg{in order to generate desired incident fields using a specific horn antenna fields,} it is useful to \sg{prepare} an approximate analytical form for the field generated by the horn. One possibility is using a Gaussian beam, which has been called \textit{quasi-optical} \cite{goldsmith_quasi-optical_1992}. A Gaussian beam is a solution to the paraxial-wave equation, which assumes a slowly-varying amplitude in the direction of propagation. In the $H$-plane ($y=0$), a Gaussian beam with TE polarization has the form \cite{saleh_fundamentals_2019}
\begin{align}
    \mathbf{E}(x,z) = A \frac{w_0}{w(z')}\exp\left[-\frac{(x')^2}{w^2(z')}-jkz'+\frac{-jk(x')^2}{2\left(z'+\frac{z_R^2}{z'}\right)}-j\arctan\left(\frac{z'}{z_R}\right)\right]\mathbf{\hat{y}}\label{eq:gaussian}
\end{align}
where $w(z')=w_0\sqrt{1+(z')^2/z_R^2}$ is the radius of the beam spot (at which the amplitude is $A/e$), $w_0$ is beam waist, $z_R=\pi w_0^2/\lambda$ is known as the Raleigh length, $k=2\pi/\lambda$, and $\lambda$ is the wavelength. Furthermore,
\begin{align}
    \begin{bmatrix}
        x'\\
        z'
    \end{bmatrix}=
    \begin{bmatrix}
        \cos\theta & -\sin\theta\\
        \sin\theta & \cos\theta
    \end{bmatrix}
    \begin{bmatrix}
        x-x_\text{bw}\\
        z-z_\text{bw}
    \end{bmatrix}
\end{align}
allows for displacing beam waist $(x_\text{bw},z_\text{bw})$ and rotating the beam by an angle $\theta$ about the beam waist.

To determine appropriate parameters to model the horn using \eqref{eq:gaussian}, a numerical fitting was performed, from which it was found that $w_0=\SI{1.00}{mm}$ and $z_\text{bw}=z_\text{ap}-\SI{4.14}{cm}$ \vtb{(obviously, $x_\text{bw}=\SI{0}{cm}$)}. In other words, the beam waist is \SI{4.14}{cm} \textit{behind} the aperture of the horn. The corresponding field is plotted in Fig.~\ref{fig:horn}b, showing a good match to the horn field and substantiating the quasi-optical approach. From Fig.~\ref{fig:horn}c, we note a discrepancy past $|x|>\SI{13.5}{cm}$. This error could be reduced by including higher-order Hermite-Gaussian modes in the model \vt{\cite{saleh_fundamentals_2019}}. \vt{Alternatively,  it is also possible to use a circular horn instead of a rectangular one, as the former provides greater coupling to the fundamental Gaussian mode \cite{wong_polarization_2015,goldsmith_quasi-optical_1992}.} However, the accuracy within the noted region for the rectangular horn is sufficient for our demonstration. Thus, a Gaussian beam provides a simple, \sg{yet} accurate model for the horn, which we will \sg{be used} to construct desired incident fields.

\section{Construction of an Arbitrary Incident Field}\label{sec:superposition}

To generate an arbitrary incident field, we may use a superposition of the fields generated by the horn for $N$ different configurations of the horn, corresponding to $N$ near field scans to be performed. In each of these experiments, the position of the beam waist $(x_{\text{bw},n},z_{\text{bw},n})$ and rotation $\theta_n$ can be adjusted, producing a different incident $\mathbf{E}_{\text{i},n}$ field and corresponding transmitted field $\mathbf{E}_{\text{t},n}$. With the surface representing a linear time-invariant (LTI) system, the fields can be superimposed with arbitrary complex weights $A_n$ to produce the fields 
\begin{align}
    \mathbf{E}_{\text{a}}(\{p_n\},x,z) = \sum_{n=1}^{N} A_n \mathbf{E}_{\text{a},n}(p_n,x,z)
\end{align}
with $\text{a}=(\text{i},\text{t})$, $p_n=\{x_{\text{bw},n},z_{\text{bw},n},\theta_n,A_n\}$ corresponding to a configuration of the horn, and $\{p_n\}$ ($n=[1,N]$) being the set of configurations. Experimentally, this can be realized by performing a near-field scan $N$ times with the given horn orientations, and subsequently numerically post-processing to find the total fields (incident or transmitted).

Now, the question is: how to select $\{p_n\}$ such that $\mathbf{E}_{\text{i}}$ is the desired incident field at the plane of the MS, which we take to be at $z=z_\text{ms}$ perpendicular to the $z$ axis? It has been shown in the literature that certain fields can be rigorously expanded as Gaussian beams. For example, in \cite{cerveny_expansion_2002} a method is presented for expanding a plane wave into a set of Gaussian beams. We however consider a simpler approach: numerical optimization. Along with simplicity, this also allows an \textit{arbitrary} incident field $\mathbf{E}_{\text{i}}(x,z_\text{ms})$ to be approximated, using an appropriate cost function.

To carry out the optimization, the cost function can be defined as
\begin{align}
    \text{cost}\left(\{p_n\}\right) = \left[\frac{1}{\Delta x}\int_{x_0}^{x_0+\Delta x} \left|\mathbf{E}_{\text{i}}\left(\{p_n\},x,z_\text{ms}\right) - \mathbf{E}_{\text{i,spec}}\left(x,z_\text{ms}\right)\right|^2\,dx\right]^{1/2}\label{eq:objective}
\end{align}
where the error from the desired field $\mathbf{E}_{\text{i,spec}}(x,z_\text{ms})$ is integrated over the region of interest where the MS is to be placed, $x=[x_0,x_0+\Delta x]$. Note that this ensures both the amplitude and phase are as desired.

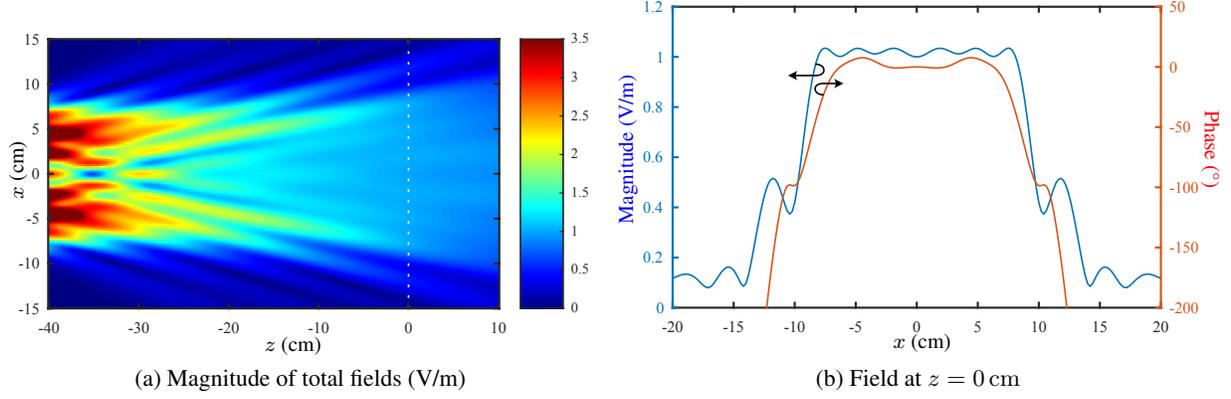
\begin{figure}[H]
    \centering
    \subcaptionbox{Magnitude of total fields (V/m)}{
        \begin{overpic}[scale=1,grid=false,trim={0 -0.2cm 0 -0.3cm},clip]{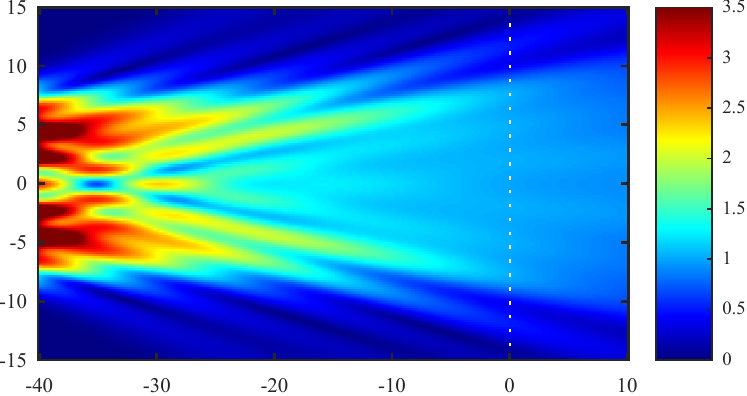}
                \put(48,0){\htext{\scriptsize{}$z$ (cm)}}
                \put(-0.5,30.5){\vtext{\scriptsize{}$x$ (cm)}}
        \end{overpic}
    }
    \subcaptionbox{Field at $z=\SI{0}{cm}$}{
        \begin{overpic}[scale=1,grid=false,trim={-0.4cm 1.1cm -0.4cm -0.3cm},clip]{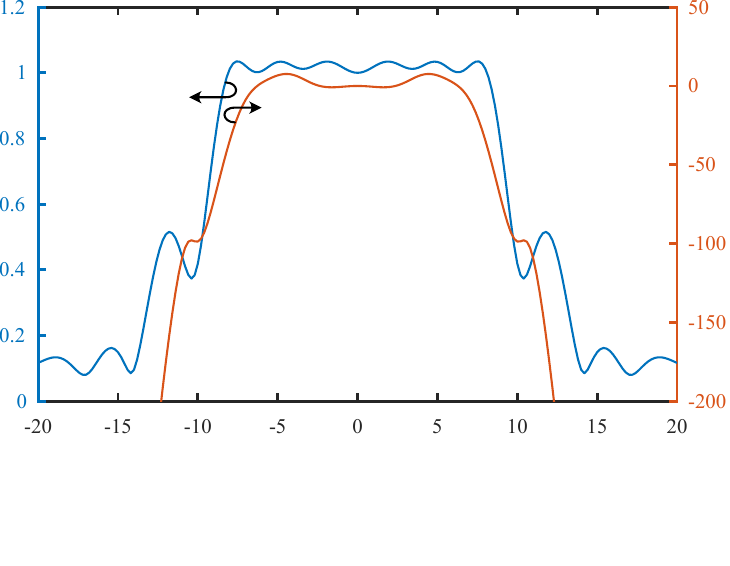}
                \put(50,0.5){\htext{\scriptsize{}$x$ (cm)}} \put(2,31.5){\vtext{\scriptsize{}\color{blue} Magnitude (V/m)}} \put(97,31.5){\vtextf{\scriptsize{}\color{red} Phase (\si{\degree})}}
        \end{overpic}
    }
    \caption{An incident field $\mathbf{E}_{\text{i}}(\{p_n\},x,z)$ generated with the superposition of $N=8$ illuminations, with the objective of uniform amplitude and phase (i.e. normally incident plane wave) over  $|x|<\SI{7.5}{cm}$.}
    \label{fig:superposition-field}
\end{figure}

As an example, we will generate $\{p_n\}$ such that the desired field has a flat phase and amplitude over $x=[-\SI{7.5}{cm},\SI{7.5}{cm}]$; i.e, a plane wave. We will set $\theta_n=\SI{0}{\degree}$, $z_\text{ms}=\SI{0}{cm}$ and $z_{\text{ap},n}=\SI{-40}{cm}$ (i.e. $z_{\text{bw},n}=\SI{-44.14}{cm}$). \vtb{In addition, the positions, $x_{\text{bw},n}$, are set at uniform intervals, resembling an antenna array.} This leaves $A_n$ as parameters for the optimization, which was performed using a genetic algorithm in MATLAB. Using $N=8$, we were able to produce the incident field in Fig.~\ref{fig:superposition-field}, using the parameters in Table~\ref{tab:superposition-params}. The amplitude has less than 4\% amplitude error and $\SI{20}{\degree}$ phase error over the specified \SI{15}{cm} span. If a better approximation is required, the number of configurations $N$ can be increased, at the cost of more near-field scans to complete. Note that while this field uses $N=8$ horn configurations, only four need to be experimentally performed if the MS is also symmetrical.

\begin{table}[H]
    \centering
    \caption{Optimized horn configurations for uniform amplitude and phase over $|x|<\SI{7.5}{cm}$, using $N=8$ illuminations.
    }\label{tab:superposition-params}
    \begin{tabularx}{0.28\columnwidth}{c c c }
        \toprule
        $n$ & $x_{\text{ap},n}$ (cm) & $A_n$     \\ \midrule
        $1,2$ & $\pm{}7.50$        & $2.39e^{-j0.63}$             \\
        $3,4$ & $\pm{}5.36$        & $5.79e^{-j0.20}$              \\
        $5,6$ & $\pm{}3.21$        & $6.90e^{j0.40}$              \\
        $7,8$ & $\pm{}1.07$        & $3.72e^{j0.45}$              \\\bottomrule
    \end{tabularx}
\end{table}

\section{Example: A Parabolic Surface}\label{sec:parabolic-example}
To demonstrate the utility of the plane wave approximation using superposition, we numerically consider the measurement of a MS synthesized for a practical field transformation. The surface is first designed with the assumption of plane wave incidence; we choose to characterize the required MS in terms of surface susceptibilities. Subsequently, we use the synthesized susceptibilities in an integral equation (IE) simulator we have developed \cite{smy_ie-gstc_2021}, with incident fields corresponding to (1) the ideal fields used for the design, (2) a single horn antenna, and (3) the superposition of horn fields determined in Section~\ref{sec:superposition}.

\begin{figure}[h]
    \centering
    \begin{overpic}[scale=1,grid=false,trim={0cm 0cm 0cm 0cm},clip]{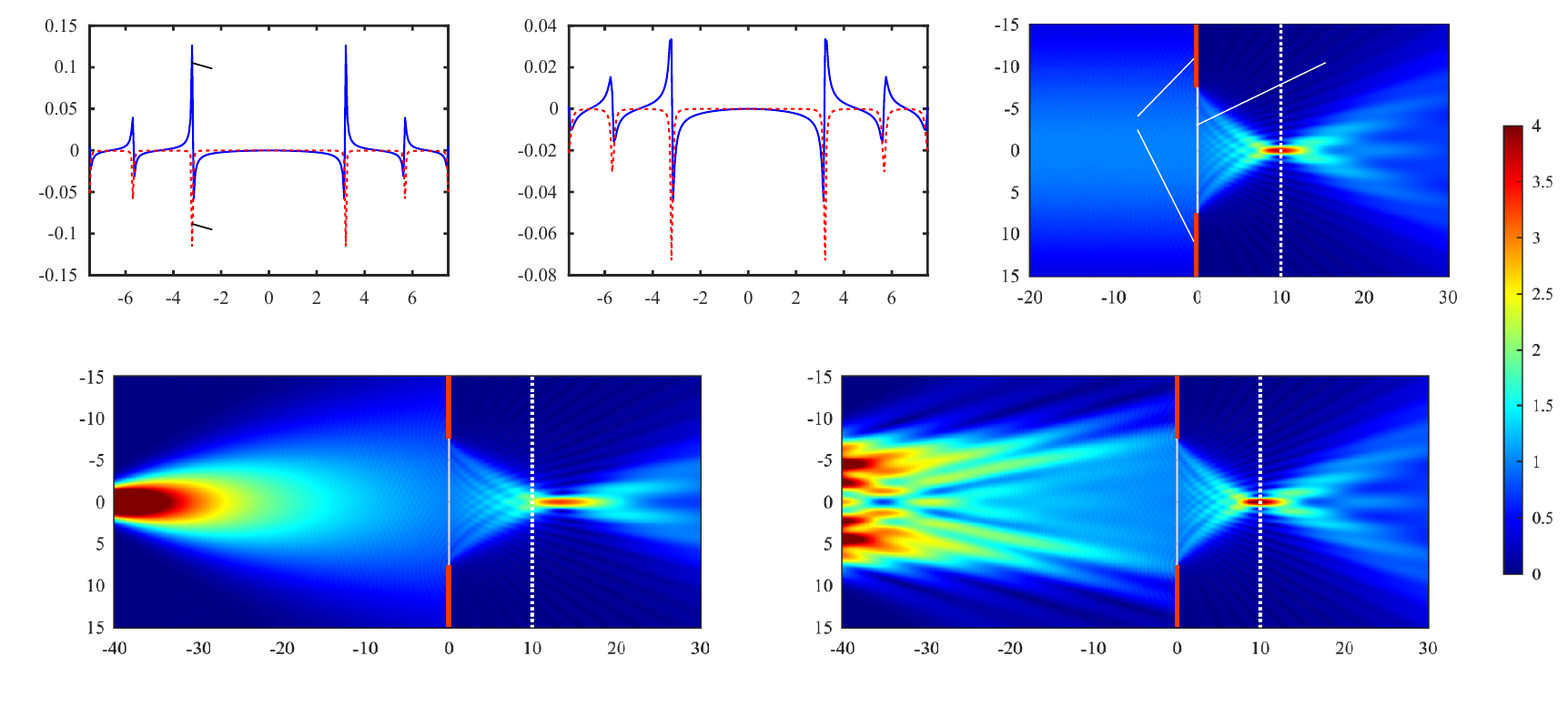}
        \put(26,0){\htext{\footnotesize{}(c) Horn illumination}}
        \put(26,2.5){\htext{\scriptsize{}$z$ (cm)}}
        \put(4.5,13.5){\vtext{\scriptsize{}$x$ (cm)}}
        \put(73,0){\htext{\footnotesize{}(d) Superposition, $N=8$}}
        \put(73,2.5){\htext{\scriptsize{}$z$ (cm)}}
        \put(51,13.5){\vtext{\scriptsize{}$x$ (cm)}}
        \put(79,23.5){\htext{\footnotesize{}(b) Ideal illumination}}
        \put(79,25.5){\htext{\scriptsize{}$z$ (cm)}}
        \put(63,35.5){\vtext{\scriptsize{}$x$ (cm)}}
        \put(69.2,37.5){\htext{\color{white}\scriptsize{}absorbing\\\color{white}\scriptsize{}boundaries}}
        \put(85,41.5){\color{white}\scriptsize{}MS}
        \put(32,23.5){\htext{\footnotesize{}(a) Synthesized susceptibilities}}
        \put(18,25){\htext{\scriptsize{}$x$ (cm)}}
        \put(47,25){\htext{\scriptsize{}$x$ (cm)}}
        \put(1.5,36){\vtext{\scriptsize{}$\Re\{\chia{ee}{yy}\}$, $\Im\{\chia{ee}{yy}\}$}}
        \put(32,36){\vtext{\scriptsize{}$\Re\{\chia{mm}{xx}\}$, $\Im\{\chia{mm}{xx}\}$}}
        \put(14.2,40.6){\htext{\scriptsize{}$\Re$}}
        \put(14.2,30.5){\htext{\scriptsize{}$\Im$}}
    \end{overpic}
    \caption{A focusing MS was designed and then illuminated with three different incident fields in (b-d).}
    \label{fig:example}
\end{figure}
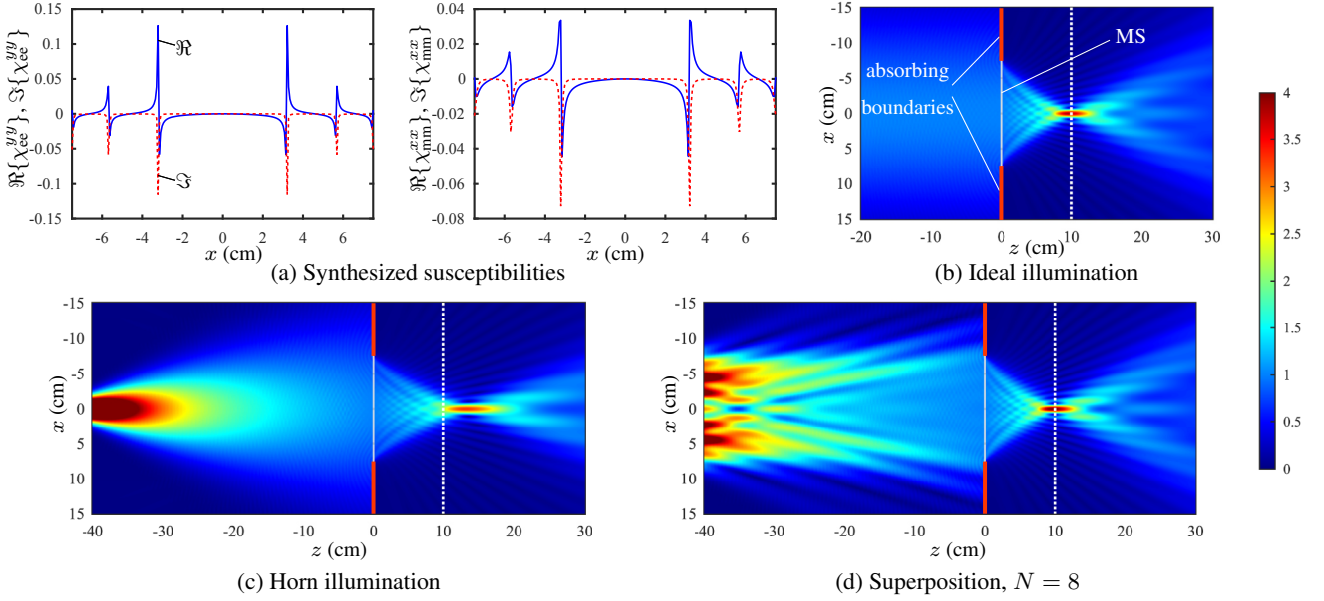

Firstly, we will design the MS to focus a normally incident plane wave to a focal point $\mathbf{r}_\text{f}=(x_\text{f},z_\text{f})$ \sg{i.e. a flat focussing lens in transmission}. We desire the transmitted field to be a cylindrical wave, given at the point $\mathbf{r}=(x,z)$ by
\begin{subequations}
\begin{align}
    E_{\text{t},y}(\mathbf{r}) &=  E_0\frac{H_{0}^{(1)}\left(k|\mathbf{r}-\mathbf{r}_\text{f}|\right)}{H_{0}^{(1)}\left(k|\mathbf{r}_\text{f}|\right)} \\
    H_{\text{t},x}(\mathbf{r}) &=E_0\frac{j(z-z_\text{f}) H_{1}^{(1)}\left(k|\mathbf{r}-\mathbf{r}_\text{f}|\right)}{\eta|\mathbf{r}|H_{0}^{(1)}\left(k|\mathbf{r}_\text{f}|\right)}\\
    H_{\text{t},z}(\mathbf{r}) &=-E_0\frac{j(x-x_\text{f}) H_{1}^{(1)}\left(k|\mathbf{r}-\mathbf{r}_\text{f}|\right)}{\eta|\mathbf{r}|H_{0}^{(1)}\left(k|\mathbf{r}_\text{f}|\right)}
\end{align}
\end{subequations}
where $H_{\left\{0,1\right\}}^{(1)}$ are Hankel functions of the first kind, representing inward propagating cylindrical waves, and the fields have been scaled such that $E_{\text{t},y}(0,0)=E_0$. Meanwhile, the incident field is a normally incident plane wave normalized such that $E_{\text{i},y}=E_0$ and $H_{\text{i},x}=-E_0/\eta$, and furthermore we desire there to be no reflection, \sg{i.e. a matched lens}. For this field transformation, suitable susceptibilities are \cite{achouri_design_2018}
\begin{subequations}
\begin{align}
    \chia{ee}{yy}(x) &= \frac{-j}{\pi \epsilon f }\left(\frac{H_{\text{t},x}(x,0)-H_{\text{i},x}(x,0)}{E_{\text{t},y}(x,0)+E_{\text{i},y}(x,0)}\right)\\
    \chia{mm}{xx}(x) &= \frac{-j}{\pi \mu f }\left(\frac{E_{\text{t},y}(x,0)-E_{\text{i},y}(x,0)}{H_{\text{t},x}(x,0)+H_{\text{i},x}(x,0)}\right)
\end{align}\label{eq:chi}
\end{subequations}
Fig.~\ref{fig:example}a shows these synthesized susceptibilities at $f=\SI{30}{GHz}$, with $(x_\text{f},z_\text{f})=(\SI{0}{cm},\SI{10}{cm})$. Note that $\Im\left\{\chia{ee}{yy}(x)\right\}<0$ and $\Im\left\{\chia{mm}{xx}(x)\right\}<0$, which indicates that the required MS is passive.

Next, the MS is illuminated using a field having $|E_\text{i}|\approx 1$ and $\angle{}E=\SI{0}{\degree}$; i.e., what the MS was designed for. \vtb{(More precisely, a Gaussian beam with $w_0=\SI{15}{cm}$ and beam waist at $z=0$ is used, with negligible error compared to a plane wave over the length of the MS.)} 
This is shown in Fig.~\ref{fig:example}b, where a focal spot at $z=\SI{10}{cm}$ is observed, \sg{as ideally desired}. This indicates that susceptibilities in Fig.~\ref{fig:example}a were correctly selected.

 However, this illumination \sg{is naturally very different from the illumination of a practical horn antenna and thus not possible to achieve in practice. The large phase curvature of the horn antenna by itself is expected to generate significantly different scattered fields as compared to the original MS designed for normally incident plane-wave, as horn fields contain large angular spectrum. To see this more clearly,} if the surface is illuminated with the horn from Section~\ref{sec:horn}, we observe the fields in Fig.~\ref{fig:example}c. Clearly, the focal spot has shifted by several centimeters, which is a significant deviation relative to the wavelength ($\lambda=\SI{1}{cm}$ at \SI{30}{GHz}). Observing this result experimentally might be (erroneously) interpreted as an indication that the MS was not designed correctly. However, as we have noted, the horn does not produce the appropriate field for characterizing the MS. \sg{This illustration thus highlights the importance of correctly choosing the Tx antenna in the measurement stages for accurate surface characterization.}

\sg{Next}, we consider the set of $N=8$ horn configurations from Table~\ref{tab:superposition-params}. After superimposing the total fields, the results in Fig.~\ref{fig:example}d is achieved, where the focal spot is \sg{now} once again observed at $z=\SI{10}{cm}$, as desired. Thus, the proposed procedure provides a good approximation of the desired incident field in this case. While we only considered $A_n$ and $x_{\text{bw},n}$ in the optimization \sg{for this specific example}, it is \sg{naturally} possible to include $z_{\text{bw},n}$ and $\theta_n$ as well, for potentially more flexibility. Furthermore, it is possible to change the desired field objective in \eqref{eq:objective} for other fields which are not necessarily plane waves.

\section{Conclusion}

Experimental metasurface characterization is limited to practical means of illumination, which at microwave frequencies is often a horn antenna. We have \sg{numerically demonstrated} a simple method of superimposing multiple near field scans using a horn antenna in different configurations, which can be used to construct incident fields which may otherwise be difficult or not possible to achieve. While this method involves additional effort by requiring multiple scans, it also provides flexibility for the incident field to be generated, simply by changing the objective of a numerical optimization which is used to find the required horn configurations for the different experiments. The method is limited to linear time-invariant MSs, which at this point in time includes most MSs that have been considered in the literature \sg{at the radio frequencies, in particular, including space-time modulated metasurfaces}. Thus, we expect that this simple procedure can be valuable for \sg{accurate} experimental characterization of metasurfaces in near-field scanning systems.


%



\section*{Acknowledgment}

The authors acknowledge funding from the Department of National Defence's Innovation for Defence Excellence and Security (IDEaS) Program in support of this work.

\ifCLASSOPTIONcaptionsoff
  \newpage
\fi



%



\bibliographystyle{IEEEtran}
\bibliography{2021_NF_Measurements,zotero}

\begin{thebibliography}{10}
\providecommand{\url}[1]{#1}
\csname url@samestyle\endcsname
\providecommand{\newblock}{\relax}
\providecommand{\bibinfo}[2]{#2}
\providecommand{\BIBentrySTDinterwordspacing}{\spaceskip=0pt\relax}
\providecommand{\BIBentryALTinterwordstretchfactor}{4}
\providecommand{\BIBentryALTinterwordspacing}{\spaceskip=\fontdimen2\font plus
\BIBentryALTinterwordstretchfactor\fontdimen3\font minus
  \fontdimen4\font\relax}
\providecommand{\BIBforeignlanguage}[2]{{%
\expandafter\ifx\csname l@#1\endcsname\relax
\typeout{** WARNING: IEEEtran.bst: No hyphenation pattern has been}%
\typeout{** loaded for the language `#1'. Using the pattern for}%
\typeout{** the default language instead.}%
\else
\language=\csname l@#1\endcsname
\fi
#2}}
\providecommand{\BIBdecl}{\relax}
\BIBdecl

\bibitem{Kuester:2003aa}
E.~F. Kuester, M.~A. Mohamed, M.~Piket-May, and C.~L. Holloway, ``Averaged
  transition conditions for electromagnetic fields at a metafilm,''
  \emph{{IEEE} Trans. Antennas Propag.}, vol.~51, no.~10, pp. 2641--2651, 2003.

\bibitem{caloz_electromagnetic_2005}
C.~Caloz and T.~Itoh, \emph{Electromagnetic {Metamaterials}: {Transmission}
  {Line} {Theory} and {Microwave} {Applications}: {The} {Engineering}
  {Approach}}.\hskip 1em plus 0.5em minus 0.4em\relax Hoboken, USA: John Wiley
  \& Sons, Inc., Nov. 2005.

\bibitem{basar_wireless_2019}
E.~Basar, M.~Di~Renzo, J.~De~Rosny, M.~Debbah, M.-S. Alouini, and R.~Zhang,
  ``Wireless {Communications} {Through} {Reconfigurable} {Intelligent}
  {Surfaces},'' \emph{IEEE Access}, vol.~7, pp. 116\,753--116\,773, 2019.

\bibitem{smy_surface_2020}
T.~J. Smy and S.~Gupta, ``Surface {Susceptibility} {Synthesis} of {Metasurface}
  {Skins}/{Holograms} for {Electromagnetic} {Camouflage}/{Illusions},''
  \emph{IEEE Access}, vol.~8, pp. 226\,866--226\,886, 2020.

\bibitem{caloz_spacetime_2020}
C.~Caloz and Z.-L. Deck-Léger, ``Spacetime {Metamaterials}—{Part} {I}:
  {General} {Concepts},'' \emph{IEEE Trans. Antennas Propag.}, vol.~68, no.~3,
  pp. 1569--1582, Mar. 2020.

\bibitem{achouri_multipolar_2021}
\BIBentryALTinterwordspacing
K.~Achouri and O.~J.~F. Martin, ``Multipolar {Modeling} of {Spatially}
  {Dispersive} {Metasurfaces},'' Mar. 2021, arXiv: 2103.10345. [Online].
  Available: \url{http://arxiv.org/abs/2103.10345}
\BIBentrySTDinterwordspacing

\bibitem{yaghjian_overview_1986}
A.~Yaghjian, ``An overview of near-field antenna measurements,'' \emph{IEEE
  Trans. Antennas Propag.}, vol.~34, no.~1, pp. 30--45, Jan. 1986.

\bibitem{gagnon_material_2002}
N.~Gagnon, J.~Shaker, P.~Berini, L.~Roy, and A.~Petosa, ``Material
  characterization using a quasi-optical measurement system,'' in \emph{Conf.
  {Dig}. {Conf}. {Precis}. {Electromagn}. {Meas}.}\hskip 1em plus 0.5em minus
  0.4em\relax IEEE, Jun. 2002, pp. 104--105.

\bibitem{wong_polarization_2015}
J.~P.~S. Wong, M.~Selvanayagam, and G.~V. Eleftheriades, ``Polarization
  {Considerations} for {Scalar} {Huygens} {Metasurfaces} and {Characterization}
  for 2-{D} {Refraction},'' \emph{IEEE Trans. Microwave Theory Techn.},
  vol.~63, no.~3, pp. 913--924, Mar. 2015.

\bibitem{selvanayagam_design_2016}
M.~Selvanayagam and G.~V. Eleftheriades, ``Design {And} {Measurement} of
  {Tensor} {Impedance} {Transmitarrays} {For} {Chiral} {Polarization}
  {Control},'' \emph{IEEE Trans. Microwave Theory Techn.}, vol.~64, no.~2, pp.
  414--428, Feb. 2016.

\bibitem{goldsmith_quasi-optical_1992}
P.~Goldsmith, ``Quasi-optical techniques,'' \emph{Proc. IEEE}, vol.~80, no.~11,
  pp. 1729--1747, Nov. 1992.

\bibitem{bruce_abcd_2006}
I.~Bruce, ``\BIBforeignlanguage{en}{{ABCD} transfer matrices and paraxial ray
  tracing for elliptic and hyperbolic lenses and mirrors},''
  \emph{\BIBforeignlanguage{en}{Eur. J. Phys.}}, vol.~27, no.~2, pp. 393--406,
  Feb. 2006, publisher: IOP Publishing.

\bibitem{smy_ie-gstc_2021}
T.~J. Smy, V.~Tiukuvaara, and S.~Gupta, ``{IE}-{GSTC} {Field} {Solver} using
  {Metasurface} {Susceptibility} {Tensors} with {Normal} {Components},'' 2021,
  arXiv: 2007.07063.

\bibitem{saleh_fundamentals_2019}
B.~E.~A. Saleh and M.~C. Teich, \emph{Fundamentals of {Photonics}},
  3rd~ed.\hskip 1em plus 0.5em minus 0.4em\relax Hoboken, USA: John Wiley \&
  Sons, Inc., 2019.

\bibitem{cerveny_expansion_2002}
V.~Červený, ``\BIBforeignlanguage{en}{Expansion of a {Plane} {Wave} into
  {Gaussian} {Beams}},'' \emph{\BIBforeignlanguage{en}{Studia Geophysica et
  Geodaetica}}, vol.~46, no.~1, pp. 43--54, Jan. 2002.

\bibitem{achouri_design_2018}
K.~Achouri and C.~Caloz, ``Design, concepts, and applications of
  electromagnetic metasurfaces,'' \emph{Nanophotonics}, vol.~7, no.~6, pp.
  1095--1116, 2018.

\end{thebibliography}

\end{document}